\documentclass[a4paper]{PoS}
\usepackage{subcaption}

\usepackage{lineno}
\title{X-ray/gamma-ray flux correlations in the BL Lac Mrk 421 using HAWC data}

\ShortTitle{X-ray/gamma-ray correlations Markarians}

\author{\speaker{J. A. Garc\'ia-Gonz\'alez}\\
        Instituto de F\'isica, UNAM\\
        E-mail: \email{jagarcia@fisica.unam.mx}}

\author{M. M. Gonz\'alez\\
        Instituto de Astronom\'ia, UNAM\\
        E-mail: \email{magda@astro.unam.mx}}

\author{N. Fraija\\
        Instituto de Astronom\'ia, UNAM\\
        E-mail: \email{nifraija@astro.unam.mx}}

\author{for the HAWC Collaboration\\
For a complete author list, see http://www.hawc-observatory.org/collaboration/}

\abstract{The HAWC gamma ray observatory is located at the Sierra Negra Volcano in Puebla, Mexico, at an altitude of 4,100 meters. HAWC is a wide field of view array of 300 water Cherenkov detectors that are continuously surveying  \textasciitilde{ }2sr of the sky, operating since March 2015. The large collected data sample allows HAWC to perform an unbiased monitoring of the BL Lac Mrk 421. This is the closest and brightest known extragalactic high-synchrotron-peaked BL Lac in the gamma-ray/X-ray bands and is extensively monitored by the Large Area Telescope (LAT) on-board the Fermi satellite, and the BAT and XRT instruments of the Swift satellite. In this work, we use 25 months of HAWC data together with Swift-XRT data to characterize potential correlations between both wavelengths.   This analysis shows that HAWC and Swift-XRT data are correlated even stronger than expected for quasi-simultaneous observations.}

\FullConference{35th International Cosmic Ray Conference - ICRC217-\\
		10-20 July, 2017\\
		Bexco, Busan, Korea}

\begin{document}

%%%%%%%%%%%%%%%%%%%%%%%%%%%%%%%%%%%%%%%%%%%
\section{Introduction}

Blazars, a subclass of active galactic nuclei (AGN), are characterized by having an outflow pointing out close to the observer's field of view.  One of the  closest, brightest and fastest varying blazars in the extragalactic X-ray/TeV sky is Markarian 421 (Mrk421). Mrk 421 located at 134 Mpc has been a frequent target of multiwavelength campaigns in order to study correlations in TeV $\gamma$-ray and X-ray bands. Observations at these energy bands have supported compelling evidence of correlated and simultaneous variability on different time scales \cite{2009ApJ...695..596H,2016ApJS..222....6B}.   In this case, different types of correlations have been reported; lineal \cite{2003ApJ...598..242A}, quadratic \cite{2008ApJ...677..906F} and fairly loose  \cite{2005ApJ...630..130B}. 
%On the other hand, high activity  detected in TeV $\gamma$-rays without its counterpart in X-rays; so-called "orphan" flares, breaks the correlation. Since there is not a definitive study about X-ray/gamma-ray correlation for Mrk 421 given that spectral variability, flare intensity, time evolution, among others are still not completely covered by the existing analysis, we believe that by performing an unbiased monitoring of this source we can contribute significantly to these correlation studies. 

The broadband spectral energy distribution (SED) of Mrk 421 exhibits a double-peaked shape; the lower energy peak is  located at X-rays and the second peak at hundreds of GeV.  Both leptonic and hadronic models have been used to model the SED of this object \cite{2011ApJ...736..131A}, implying comparable jet powers.

%In the Leptonic scenario, a one-zone synchrotron self-Compton (SSC) model has been required \cite{2011ApJ...736..131A}.   In the hadronic scenario, the first peak is interpreted by electron synchrotron radiation and the second peak by hadronic interactions \cite{2011ApJ...736..131A, 2015APh....70...54F}. The TeV $\gamma$-ray and X-ray correlation is usually interpreted in the SSC model \cite{Aleksic2015}.  In this framework, the emitting region moving at ultra-relativistic velocities in a collimated jet has an electron distribution. These  Fermi-accelerated electrons injected in the emitting region and confined to it by a magnetic field, radiate photons by synchrotron emission and after that scatter them up to higher energies by inverse Compton.

In this paper we are presenting correlation studies of X-ray/gamma-ray bands using 25 months of data from November 26th, 2014 to December 31st, 2016.  The work is organized as follows. In Section~\ref{sec:datasets} we are describing the data set that is used to calculate the X-ray/gamma-ray correlation for Mrk 421. Section~\ref{sec:AgostiniDef} presents the analysis and results and in Section~\ref{sec:theory} we discuss some theoretical implications. 

%Section~\ref{sec:AgostiniDef} presents the method used to calculate the correlation obtained in sec~\ref{sec:corr}. In section~\ref{sec:theory} we derive some theoretical implications that can be obtained from correlation studies. Finally we summarize our conclusions in section~\ref{sec:conclusions}

%%%%%%%%%%%%%%%%%%%%%%%%%%%%%%%%%%%%%%%%%
\section{\label{sec:datasets}Data Sets}

\begin{figure}[ht]
	\begin{center}
     \includegraphics[width=.45\textwidth]{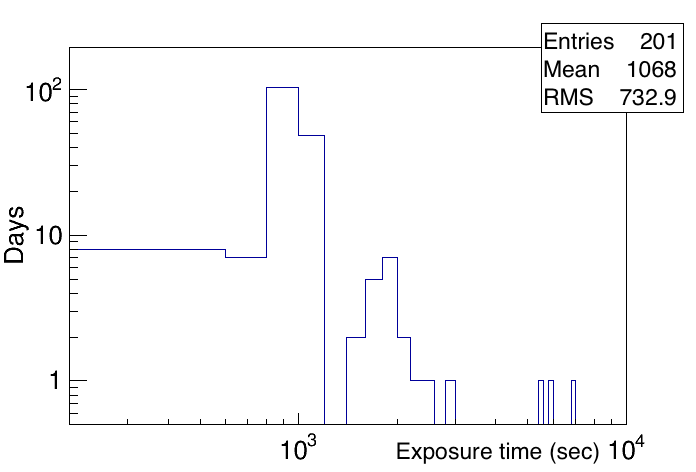}
     \includegraphics[width=.45\textwidth]{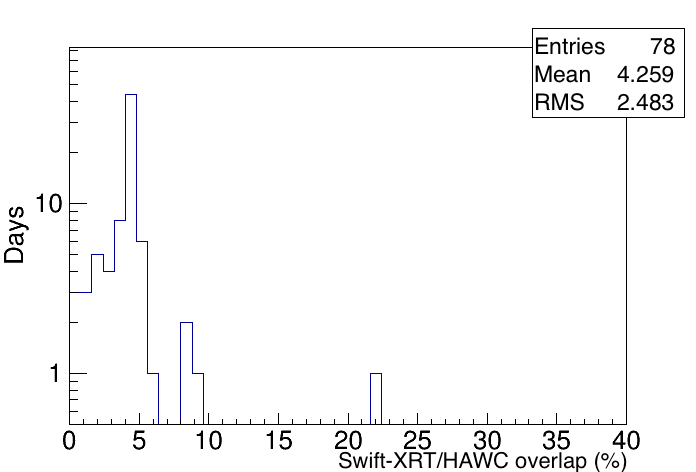}
     \caption{Left panel: Swift-XRT daily exposure time during HAWC observations period. Right panel: Fraction of exposure times when requiring simultaneity of SWIFT/XRT observation with HAWC.}
     \label{timeWINDOW}
     \end{center}
\end{figure}

The HAWC daily light curve for Mrk 421 during the period from November 2014 to April 2016 is published in~\cite{LCRobert}. We use the same analysis to extend our data set up to December 2016. For X-ray data, we use public Swift-XRT light curves in photon counts from the web site given by ~\cite{XRTData}. The time bin size depends on the count rate. In particular during the period of HAWC observations, the Swift-XRT count rate was in general below $10$ cts/s. The minimum time bin size was such that a minimum of $200$ counts per bin were collected during a mean exposure time of $1068$ s as observed in left panel of Figure ~\ref{timeWINDOW}, while the transit of Mrk 421 over HAWC last in average \textasciitilde{ }6 hr per day. The mean fraction of exposure times is 4.7$\%$.   

We extract the data for those 110 days when both instruments observed Mrk 421 and combine them per day and week. The mean fraction of exposure times slightly changes to 4.3$\%$ when requiring simultaneity of SWIFT/XRT observation with HAWC as shown in right panel of Figure ~\ref{timeWINDOW}. There are not simultaneous observations for 32 of the 110 (29$\%$) days included in this analysis.

%%%%%%%%%%%%%%%%%%%%%%%%%%%%%%%%%%%%%%%%%
%\section{\label{sec:AgostiniDef}Estimation of the correlation}
\section{\label{sec:AgostiniDef}Analysis and Results}

We use the maximum likelihood approach discussed by D'Agostini~\cite{DAgostini} to determine the correlation between SWIFT-RXT and HAWC data. This approach assumes that both data sets are linearly correlated with an intrinsic scatter $\sigma$ or,

\begin{equation}
F_{\gamma} = aF_{x} + b
\end{equation}

\noindent where $F_{\gamma}$ is the VHE integrated flux and $F_{x}$ is the X-ray count rate. To know more about the explicit role of the parameter $\sigma$ please refer to ~\cite{DAgostini}.
% Then, the parameter values can be estimated by minimizing the minus-log-likelihood function given by D'Agostini~\cite{DAgostini}. We have obtained the values for the parameters given en Table~\ref{tab:fitpar}. The fits are shown in Figure ~\ref{CorrMrk421} and, green, blue and yellow lines represent 1, 2 and 3 $\sigma$ scatter around the best fit, respectively. Although the uncertainties in the parameters are high for both cases, daily and weekly, the Pearson coefficients and the p-values indicate a strong correlation between HAWC and SWIFT-XRT observations. Also, the values obtained for the slope $a$ are not consistent with a flat fit, while the values for $b$  are consistent with zero. Thus, we have fixed the value of $b$ to zero in order to obtain a better estimation of the other parameters. The results are given in Table ~\ref{tab:fitpar}. We observed that the scatter, quantified by $\sigma$, for the case of daily fluxes is higher than the expected by correcting the obtained for weekly fluxes. In the most optimistic case, the correction should scale by $\sqrt {7}$. This may indicate an extra daily variability in the fluxes that is intrinsic to the source. The measurement of highest flux of  $110 \times 10^{-12}$ ph cm$^{-2}$ s $^{-1}$ supports this assumption.

\begin{figure}[ht]
    \begin{subfigure}{0.5\textwidth}
     \includegraphics[width=1.0\textwidth]{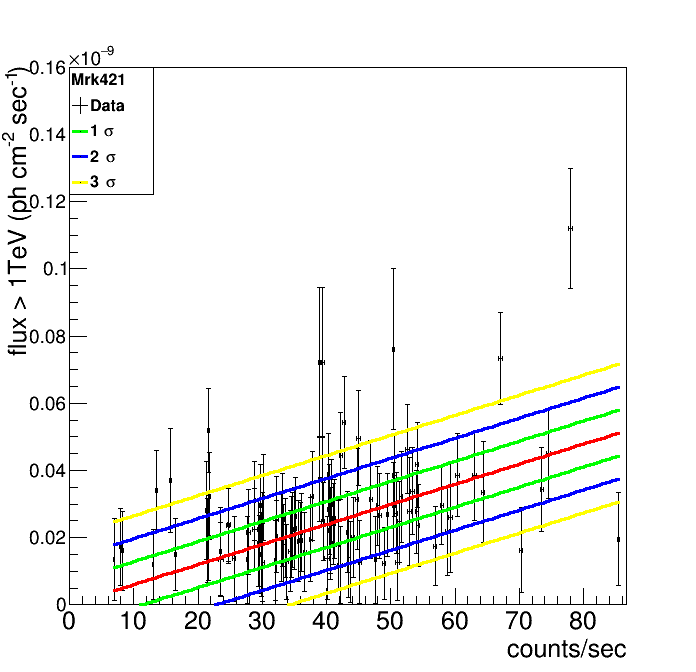}
     \caption{}
     \label{CorrDaily110Mrk421}
     \end{subfigure}
     \begin{subfigure}{0.5\textwidth}
     \includegraphics[width=1.0\textwidth]{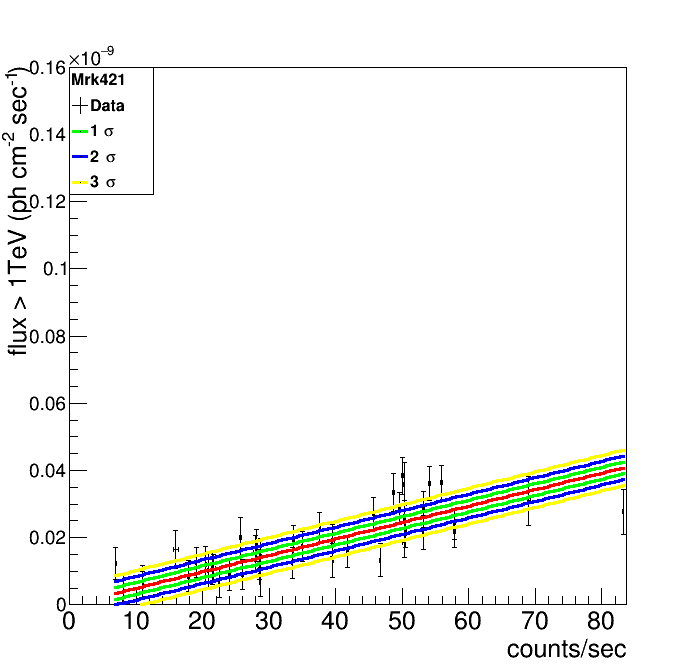}
     \caption{}
     \label{CorrWeek44Mrk421}
     \end{subfigure}
    \caption{X-ray/gamma-ray correlation of Mrk421 for daily binning (a) with 110 data points and weekly binning (b) with 45 data points. The total data sample used cover 25 months of HAWC data that has some level of overlap with Swift-XRT data. Integrated flux for weekly/daily average for HAWC (Y axis) and X-ray rate for Swift-XRT. Colored lines represent 1,2 and 3 $\sigma$ parameter described in section~\ref{sec:AgostiniDef}. Fit values for parameters are summarized in Table~\ref{tab:fitpar}.  }
    \label{CorrMrk421}
\end{figure}

Then, the parameter values can be estimated by minimizing the minus-log-likelihood function given by D'Agostini~\cite{DAgostini}. We have obtained the values for the parameters given in Table~\ref{tab:fitpar}. The fits are shown in Figure ~\ref{CorrMrk421} and, green, blue and yellow lines represent 1, 2 and 3 $\sigma$ scatter around the best fit, respectively. Although the uncertainties in the parameters are high for both cases, daily and weekly, the Pearson coefficients and the p-values indicate a strong correlation, within 3$\sigma$, between HAWC and SWIFT-XRT observations. The values obtained for the slope $a$ are not consistent with a flat fit, while the values for $b$  are consistent with zero. Thus, we have fixed the value of $b$ to zero in order to obtain a better estimation of the remaining parameters. The results are given in Table ~\ref{tab:fitpar}. Both fits are consistent. However, the scatter, quantified by $\sigma$, for the case of daily fluxes is higher than the expected by correcting the obtained for weekly fluxes. In the most optimistic case of having all the measurements per day, the correction should scale by $\sqrt {7}$. A possible explanation is an extra daily variability intrinsic to the source. If this is the case, it also explains the measurement of the highest flux of  $110 \times 10^{-12}$ ph cm$^{-2}$ s $^{-1}$ in daily time scales inconsistent with the best fit for more than 6$\sigma$.

\begin{table}
\begin{center}
     \begin{tabular}{l|c|c|c|c|c}
     \hline
     binning & $a (\times 10^{-13})$ & $b (\times 10^{-12})$ & $\sigma (\times 10^{-12})$ & Pearson Coeff. & p-value  \\
     \hline
  daily     & $5.62 \pm 4.25$ & $1.58 \pm 3.29$ & $6.90 \pm 9.35$  & 0.483 & $4.54\times10^{-8}$\\
  weekly & $4.39 \pm 2.37$ & $2.11 \pm 3.55$ & $1.65 \pm 1.13$  & 0.788 & $6.63\times10^{-11}$\\     
  \hline
  daily     & $5.97 \pm 1.62$     & 0 fixed & $6.84 \pm 8.4$  & 0.483 & $4.54\times10^{-8}$\\
  weekly & $4.89 \pm 0.98$     & 0 fixed & $1.76 \pm 6.2$  & 0.788 & $6.63\times10^{-11}$\\     
     \end{tabular}   
\end{center}
\caption{Parameters obtained from the Likelihood fit using the D'Agostini method described in section~\ref{sec:AgostiniDef}. First row shows the fit values for daily binning and second row for weekly binning.}
\label{tab:fitpar}
\end{table}

\section{\label{sec:theory}Theoretical Interpretation}
The broadband SED of Mkr421 is usually interpreted in the SSC and hadronic models. Electrons accelerated and confined in the emitting region by the magnetic field radiate by synchrotron emission peaking at X-rays. These synchrotron photons, in the leptonic scenario,  are scattered up to higher energies only by inverse Compton and in the hadronic model serve as targets for the pion productions, producing high energy photons by $\pi^0$ decay products, $\pi^\pm$ cascade and $\mu$ synchrotron radiation among others. Therefore, an X-ray/VHE gamma-ray lineal correlation is expected only in the SSC scenario.\\
In this framework, the synchrotron spectrum is given by,
{\small
\begin{eqnarray}
\label{eq:espsyn}
&&F^{\rm syn}_\nu\propto\cases{ 
 (\frac{\epsilon^{\rm syn}_\gamma}{\epsilon_{\gamma,m}})^{\frac{-(\alpha-3)}{2}}  &  $\epsilon^{\rm syn}_{\gamma,m} < \epsilon^{\rm syn}_\gamma < \epsilon^{\rm syn}_{\gamma,c}$,\cr
(\frac{\epsilon^{\rm syn}_{\gamma,c}}{\epsilon^{\rm syn}_{\gamma,m}})^{\frac{-(\alpha-3)}{2}}    (\frac{\epsilon^{\rm syn}_\gamma}{\epsilon^{\rm syn}_{\gamma,c}})^{\frac{-(\alpha-2)}{2}},           &  $\epsilon^{\rm syn}_{\gamma,c} < \epsilon^{\rm syn}_\gamma $\,,\cr
}
\end{eqnarray}
}
where the spectral synchrotron breaks $\epsilon^{\rm syn}_{\rm \gamma,m}$ and $\epsilon^{\rm syn}_{\rm \gamma,c}$ are given as,
\begin{eqnarray}\label{synrad}
\epsilon^{\rm syn}_{\rm \gamma,m} &\propto& \Gamma\,U_B^{1/2}\,  N_e^{-2},\cr
\epsilon^{\rm syn}_{\rm \gamma,c} &\propto& \Gamma^3\, U_B^{-3/2}\, r_d^{-2}.
\end{eqnarray}
Here, $\Gamma$ is the bulk Lorentz factor, $r_d$ is the size of the emitting radio, $U_B=B^2/8\pi$ is the magnetic density and $N_e$ is the density of electrons. The inverse Compton scattering of synchrotron spectrum can be written as 
{\small
\begin{eqnarray}
\label{eq:espic}
&&F^{\rm ssc}_\nu\propto\cases{ 
 (\frac{\epsilon^{\rm ssc}_\gamma}{\epsilon^{\rm ssc}_{\gamma,m}})^{\frac{-(\alpha-3)}{2}}  &  $\epsilon^{\rm ssc}_{\gamma,m} < \epsilon^{\rm ssc}_\gamma < \epsilon^{\rm ssc}_{\rm \gamma,c}$,\cr
(\frac{\epsilon^{\rm ssc}_{\gamma,c}}{\epsilon^{\rm ssc}_{\gamma,m}})^{\frac{-(\alpha-3)}{2}}    (\frac{\epsilon^{\rm ssc}_\gamma}{\epsilon^{\rm ssc}_{\rm \gamma,c}})^{-(\alpha-2)/2},           &  $\epsilon^{\rm ssc}_{\rm \gamma,c} < \epsilon^{\rm ssc}_\gamma  $\,,\cr
}
\end{eqnarray}
}
where the characteristic ($\epsilon^{\rm ssc}_{\rm \gamma,m}$) and cooling ($\epsilon^{\rm ssc}_{\rm \gamma,c}$) SSC energies are
\begin{eqnarray}\label{icrad}
\epsilon^{\rm ssc}_{\rm \gamma,m} &\propto& \Gamma\,U_B^{1/2}\,  N_e^{-4},\cr
\epsilon^{\rm ssc}_{\rm \gamma,c} &\propto&\,\Gamma^5\, U_B^{-7/2}\, r_d^{-4}\,.
\end{eqnarray}
From eqs. (\ref{synrad}), (\ref{eq:espsyn}), (\ref{icrad}) and (\ref{eq:espic}), we can show that VHE $\gamma$-ray and X-ray fluxes are linearly correlated . Thus, it can be written as, 

\begin{equation}
F^{\rm ssc}_\nu=\mathcal{F}(\Gamma,\,B,\, N_e,\, r_d)\,F^{\rm syn}_\nu\,,
\end{equation}
%
%the first peak at X-rays is explained by electron synchrotron radiation and the second peak that extends from low-energy $\gamma$-rays to VHE $\gamma$-rays, the synchrotron proton blazar (SPB) model is used.
%
where $\mathcal{F}$ is a function that depends on the bulk Lorentz factor, the size of the emitting radio, the strength of magnetic field and the density of electrons.   From the previous quantities, we observe that the density of radiating electrons is the only quantity that varies in time, so the electrons are cooled as they continually emit synchrotron photons.  Therefore, number of radiating electrons  could be responsible of the small differences between daily and weekly correlations. Similarly, fluctuations of the magnetic field in small timescales could affect correlations in different timescales. However, we have not observed significant differences between daily and weekly correlations. More data will be included as HAWC keeps operations.

%%%%%%%%%%%%%%%%%%%%%%%%%%%%%%%%%%%%%%%bibliography

\acknowledgments
We acknowledge the support from: the US National Science Foundation (NSF); the
US Department of Energy Office of High-Energy Physics; the Laboratory Directed
Research and Development (LDRD) program of Los Alamos National Laboratory;
Consejo Nacional de Ciencia y Tecnolog\'{\i}a (CONACyT), M{\'e}xico (grants
271051, 232656, 260378, 179588, 239762, 254964, 271737, 258865, 243290,
132197), Laboratorio Nacional HAWC de rayos gamma; L'OREAL Fellowship for
Women in Science 2014; Red HAWC, M{\'e}xico; DGAPA-UNAM (grants IG100317,
IN111315, IN111716-3, IA102715, 109916, IA102917); VIEP-BUAP; PIFI 2012, 2013,
PROFOCIE 2014, 2015;the University of Wisconsin Alumni Research Foundation;
the Institute of Geophysics, Planetary Physics, and Signatures at Los Alamos
National Laboratory; Polish Science Centre grant DEC-2014/13/B/ST9/945;
Coordinaci{\'o}n de la Investigaci{\'o}n Cient\'{\i}fica de la Universidad
Michoacana. Thanks to Luciano D\'{\i}az and Eduardo Murrieta for technical support.
\end{document}